\pdfoutput=1
\documentclass[sigconf]{acmart}

\AtBeginDocument{%
  \providecommand\BibTeX{{%
    \normalfont B\kern-0.5em{\scshape i\kern-0.25em b}\kern-0.8em\TeX}}}

\acmConference[ICSE 2024]{46th International Conference on Software Engineering}{April 2024}{Lisbon, Portugal}
\usepackage{balance}
\usepackage{caption}
\usepackage{subcaption}
\usepackage{subfiles}
\usepackage{enumerate}
\usepackage{xcolor}
\usepackage{bm}
\usepackage{listings}
\usepackage{amsmath}
\usepackage{pifont}
\usepackage{hyperref}
\usepackage{cleveref}
\usepackage{multirow}
\usepackage[ruled,linesnumbered]{algorithm2e}
\usepackage[normalem]{ulem}

\usepackage{comment}
\usepackage{dutchcal}
\usepackage{threeparttable}
\usepackage{listings}
\usepackage{mathtools}
\usepackage{enumitem}

\usepackage{flushend}

\crefformat{section}{\S#2#1#3}
\crefname{figure}{Figure}{Figures}
\crefname{table}{Table}{Tables}
\crefname{algorithm}{Algorithm}{Algorithms}
\usepackage{amssymb}
\usepackage{diagbox}
\usepackage{mathrsfs}

\newcommand{\wenbo}[1]{\textcolor{green}{}}
\newcommand{\xingyu}[1]{\textcolor{orange}{}}
\newcommand{\ruofan}[1]{\textcolor{blue}{}}
\newcommand{\loancold}[1]{\textcolor{purple}{}}
\newcommand{\cw}[1]{\textcolor{teal}{}}

\newcommand{\name}{ModuleGuard\xspace}
\newcommand{\extractor}{InstSimulator\xspace}
\newcommand{\resolver}{EnvResolution\xspace} 
\newcommand{\pip}{\code{pip}\xspace}
\newcommand{\code}[1]{\texttt{{\hyphenchar \font=\defaulthyphenchar \small \detokenize{#1}}}}

\copyrightyear{2024}
\acmYear{2024}
\setcopyright{acmlicensed}\acmConference[ICSE '24]{2024 IEEE/ACM 46th International Conference on Software Engineering}{April 14--20, 2024}{Lisbon, Portugal}
\acmBooktitle{2024 IEEE/ACM 46th International Conference on Software Engineering (ICSE '24), April 14--20, 2024, Lisbon, Portugal}
\acmDOI{10.1145/3597503.3639221}
\acmISBN{979-8-4007-0217-4/24/04}

\def\BibTeX{{\rm B\kern-.05em{\sc i\kern-.025em b}\kern-.08em
    T\kern-.1667em\lower.7ex\hbox{E}\kern-.125emX}}

\begin{document}
\title{\name: Understanding and Detecting Module Conflicts in Python Ecosystem}

\author{Ruofan Zhu}
\email{zhuruofan@zju.edu.cn}
\orcid{0009-0005-5181-4797}
\affiliation{%
  \institution{Zhejiang University}
  % \streetaddress{P.O. Box 1212}
  \city{Hangzhou}
  \country{China}
  \postcode{310000}
}

\author{Xingyu Wang}
\email{wangxingyu@zju.edu.cn}
\orcid{0009-0009-9988-8065}
\affiliation{%
  \institution{Zhejiang University}
  % \streetaddress{P.O. Box 1212}
  \city{Hangzhou}
  \country{China}
  \postcode{310000}
}

\author{Chengwei Liu}
\email{chengwei001@e.ntu.edu.sg}
\orcid{0000-0003-1175-2753}
\affiliation{%
  \institution{Nanyang Technological University}
  % \streetaddress{P.O. Box 1212}
  % \city{Hangzhou}
  \country{Singapore}
  % \postcode{310000}
}

\author{Zhengzi Xu}
\email{zhengzi.xu@ntu.edu.sg}
\orcid{0000-0002-8390-7518}
\affiliation{%
  \institution{Nanyang Technological University}
  % \streetaddress{P.O. Box 1212}
  % \city{Hangzhou}
  \country{Singapore}
  % \postcode{310000}
}

\author{Wenbo Shen}
\authornote{Wenbo Shen is the corresponding author.}
\email{shenwenbo@zju.edu.cn}
\orcid{0000-0003-2899-6121}
\affiliation{%
  \institution{Zhejiang University}
  % \streetaddress{P.O. Box 1212}
  \city{Hangzhou}
  \country{China}
  \postcode{310000}
}

\author{Rui Chang}
\email{crix1021@zju.edu.cn}
\orcid{0000-0002-0178-0171}
\affiliation{%
  \institution{Zhejiang University}
  % \streetaddress{P.O. Box 1212}
  \city{Hangzhou}
  \country{China}
  \postcode{310000}
}

\author{Liu Yang}
\email{yangliu@ntu.edu.sg}
\orcid{0000-0001-7300-9215}
\affiliation{%
  \institution{Nanyang Technological University}
  % \streetaddress{P.O. Box 1212}
  % \city{Hangzhou}
  \country{Singapore}
  % \postcode{310000}
}

% \author{Kui Ren}
% \email{kuiren@zju.edu.cn}
% \orcid{0000-0002-1969-2591}
% \affiliation{%
%   \institution{Zhejiang University}
%   % \streetaddress{P.O. Box 1212}
%   \city{Hangzhou}
%   % \state{Ohio}
%   \country{China}
%   \postcode{310000}
% }

\begin{abstract}
Python has become one of the most popular programming languages for software development due to its simplicity, readability, and versatility. As the Python ecosystem grows, developers face increasing challenges in avoiding module conflicts, which occur when different packages have the same namespace modules. Unfortunately, existing work has neither investigated the module conflict comprehensively nor provided tools to detect the conflict. 
Therefore, this paper systematically investigates the module conflict problem and its impact on the Python ecosystem. 
We propose a novel technique called \extractor, which leverages semantics and installation simulation to achieve accurate and efficient module extraction. Based on this, we implement a tool called \name to detect module conflicts for the Python ecosystem. 

For the study, we first collect 97 MC issues, classify the characteristics and causes of these MC issues, summarize three different conflict patterns, and analyze their potential threats. Then, we conducted a large-scale analysis of the whole PyPI ecosystem (4.2 million packages) and GitHub popular projects (3,711 projects) to detect each MC pattern and analyze their potential impact. We discovered that module conflicts still impact numerous TPLs and GitHub projects. 
This is primarily due to developers' lack of understanding of the modules within their direct dependencies, not to mention the modules of the transitive dependencies. 
Our work reveals Python's shortcomings in handling naming conflicts and provides a tool and guidelines for developers to detect conflicts.
\end{abstract}



\maketitle

%At the time of submission, all papers must not exceed 10 pages for all text and figures plus 2 pages for references.res

\section{Introduction}
\label{sec:introduction}
Namespace conflicts are a ubiquitous challenge in the field of computer science. However, with the thriving development of the software supply chain in recent years, there has been explosive growth in the number of open-source software, making the issue of namespace conflicts more pressing~\cite{Namingcollision}. According to the Sonatype report~\cite{sonatype2021}, the number of open-source software represents a 20\% year-over-year growth globally. 
As software systems become larger and more diverse, the likelihood of encountering namespace conflicts increases significantly.

While namespace conflicts have long been a topic of discussion, different language ecosystems have taken different approaches to address namespace conflict issues. For instance, the Java ecosystem uses groupId, artifactId, and version to name an open-source package uniquely and extracts the package’s modules to a unique path formed by this triplet~\cite{maven2021}. Rust isolates different open-source packages by placing them in separate folders~\cite{cargo}. Similarly, Python Package Index (PyPI) uses the project name as a unique identifier for an open-source package, but the module names within the package are not unique~\cite{pip}. As a result, we have found that it poses some threats in handling naming conflicts. First, unlike other languages, Python does not isolate each package that has been downloaded locally but instead installs them together by default. This results in an impact of module overwriting when two packages with conflicting module namespaces are installed simultaneously. 

Second, as an interpreted language, Python determines which specific module to import at runtime, unlike compiled languages that can report errors in advance. Hence, when importing packages at runtime, the existence of modules in different paths may lead to conflicts, posing a threat of import confusion. We refer to the conflicts caused by module namespaces as \underline{M}odule \underline{C}onflicts (MC). These two threats of module conflicts are unique to the Python ecosystem since they are caused by Python’s specific mechanisms for handling namespace conflicts. These threats can cause the program to break locally existing third-party packages when installed, causing environmental damage that is difficult to repair. They can cause function errors in program execution, leading to bugs that are hard to trace and fix. Moreover, the MC issues can inhibit the normal update of packages, breaking their integrity.

Existing work provides in-depth research and analysis on a number of issues specific to the Python ecosystem. Cheng et al.~\cite{cheng2022conflict} propose PyCRE, a new approach to automatically infer Python-compatible runtime environments with domain knowledge graphs. Ye et al.~\cite{ye2022knowledge} propose PyEGo, a knowledge-based technique that can automatically infer dependencies' compatible versions for Python programs. Unfortunately, when dealing with module conflicts, they both simply consider the conflicting modules belonging to the most popular packages. Wang et al.~\cite{wang2021restoring} present SnifferDog, an approach that can restore the execution environments of Jupyter notebooks. Garcia et al.~\cite{horton2019dockerizeme} present DockerizeMe, a tool that can generate Dockerfiles for Python projects.
However, when dealing with module conflicts, they all use a limited number of module-to-package mappings, which do not realize conflicts.  Other works primarily focus on dependency conflicts~\cite{wang2020watchman, cheng2022conflict, Wang2022SmartPip, lieasypip} that resolve different software components require incompatible versions of the same dependency, dependency diagnosis~\cite{cao2022towards} that fix incorrect dependency configurations, and detecting malicious packages~\cite{alfadel2021empirical, ruohonen2021large, vu2020typosquatting, duan2020towards, guo2023empirical} for PyPI packages and dependencies. These works do not systematically address the MC problem or detect MC. They either ignore the existence of multiple packages that provide the same module or rely on incomplete module-to-package mappings that can not cover all possible scenarios.

To fill this gap, this paper conducts a systematic study to investigate the module conflict problem and its impact on the Python ecosystem. To achieve the large-scale study, we first propose a novel technique named \textit{installation-free module extraction} (\textit{\extractor} in short). It extracts the semantics of different configuration files from multiple dimensions and then simulates the installation process according to the semantics to extract exact modules. Our evaluation shows that \extractor achieves over 95\% accuracy and can complete the module extraction of all four million packages in 10 hours with 40 threads in parallel. Then, based on \extractor we implement a novel tool named \name for the study. 

To conduct the study, we first collect 97 MC issues on GitHub and StackOverflow, classify the characteristics and causes of these MC issues, summarize three different conflict patterns, and analyze their potential threats. Next, using \name, we conduct a large-scale analysis of the whole PyPI ecosystem (4.2 million packages) and GitHub popular projects (3,711 projects) to detect each MC pattern and analyze their potential impact. We have discovered that there are still numerous open-source software packages in PyPI that are impacted by module conflicts among their dependencies. This is primarily due to developers lacking understanding of the modules within their direct and indirect dependencies.
Our work not only reveals Python's shortcomings in handling naming conflicts but also provides tools and guidelines for developers to detect conflicts when developing and debugging.

To summarize, this paper makes the following contributions.
\begin{itemize}[leftmargin=*]
\item \textbf{New study.} We conduct a systematic study on module conflicts (MC) in the Python ecosystem. 
We conducted an issue study from GitHub and StackOverflow and summarized three MC patterns- module-to-TPL, module-to-Lib, and module-in-Dep conflicts and their two potential threats. 

\item \textbf{New technique.} We propose \extractor, which leverages the semantics and installation simulation to achieve accurate and efficient module extraction. 
Based on this, we implement a tool \name to detect MCs for the Python ecosystem. We construct benchmarks for evaluating the capabilities of module information extraction and dependency graph resolution.
% , and experimental results demonstrate that \name is more than 95\% and 96\% accurate in these two aspects, respectively.

\item \textbf{Ecosystem-scale analysis.} Utilizing \name, we conduct a large-scale study and analyze 4.2 million packages on PyPI (434,823 latest version packages as of April 2023). We get a lot of interesting findings, shed some light on the nature of module conflicts, and provide some guidance for developers.

\item \textbf{Issue reporting.} We examine 93,487 tags of 3,711 popular GitHub projects, of which 108 are now or ever affected by MC. We reported issues, and a lot of issues have been confirmed and fixed. This proves that our work can help developers understand previously unrealized errors and help them fix potential threats.
\end{itemize}

\ruofan{final fix}This paper is organized as follows.
We introduce background knowledge in~\cref{sec:background}. In~\cref{sec:moduleguard}, we propose our \name tool and introduce it in detail. In ~\cref{sec:evaluation} we evaluate \name in different metrics. In ~\cref{sec:empirical-study} we conduct an ecosystem-scale study on MC, including issues study, GitHub projects, and PyPI packages. ~\cref{sec:limitations} and \cref{sec:disscussion} present some limitations and discussion. Related work is described in~\cref{sec:related-work}. Finally, we conclude the whole paper in~\cref{sec:conclusion}.

\section{Background}
\label{sec:background}
In this section, we provide the necessary background for our study. We first explain how Python code is managed and how Python modules are shared among developers. Then, we describe how PyPI handles third-party libraries (TPLs) and their dependencies. Finally, we illustrate two examples of module conflicts (MCs), which are the main problem we address in this paper.

\textbf{Python code management.}
Python uses \pip~\cite{pip} as its official package manager for downloading and installing TPLs from PyPI. When \pip receives a package request, it first resolves the constraint and selects a suitable version of the package. Then, it downloads the package to a local temporary folder. If the package is a source distribution package, \pip extracts all the files, compiles the package, generates metadata files, and installs the package based on the metadata files. If the package is a binary distribution package, \pip installs it directly based on the metadata files embedded in it. By default, \pip installs third-party packages into the \code{site-packages} folder unless the user specifies a different folder using the \code{target (-t)} argument.

After installing packages, users can import modules from them in their Python codes. The import process consists of two steps: first, the user declares the name of the module to be imported (e.g., \code{import bs4}). Second, the Python interpreter searches for the module when it executes the import statement. The Python interpreter first looks in the module cache that records the modules already imported. If not found in the cache, the interpreter searches in the \code{sys.path} order, which is a list of directories where Python looks for modules. Once the interpreter finds a module with the same name as the import statement, it stops the search and returns the module.

% When \pip downloads third-party packages from PyPI, it will resolve the constraint and select a suitable version package. If the selected package is a source distribution package, \pip downloads it to a local temporary folder, extracts all the files, compiles the package while generating metadata files, and installs the package based on the metadata files. On the other hand, if the selected package is a binary distribution package, \pip downloads it and installs it directly based on the metadata files in it since it was built when it was packaged. In default, \pip downloads third-party packages into the \textit{site-packages} folder or into a specific folder if users specify the \textit{target (-t)} argument.

% After downloading packages, users can import modules from them. It consists of two steps, first the user declares the module's name to be imported (e.g. \code{import bs4}). Then the Python interpreter searches for the module when it runs that import statement. the Python interpreter first looks in the module cache that records the modules already imported, then searches in \code{sys.path} order, and once it finds a module named after the statement, it returns directly without continuing the search.

\textbf{Python dependency management.}
\label{sec:Python-dependency-management}
Python developers can specify the dependencies of their projects in \code{requirement.txt} and use the command \code{pip install -r requirement.txt} to install them. Alternatively, developers can package and upload their projects to PyPI after declaring the dependencies in a configuration file, such as \code{setup.py} or \code{pyproject.toml}. The configuration file is compiled to generate metadata files, which contain the dependency information. The dependency information is stored in the \code{requires.txt} file of the source distribution package or in the \code{METADATA} file of the binary distribution package.

Python supports various types of dependencies, such as build, install, extra, and test dependencies. Each type can also be conditional on the local environment so that \pip will install them selectively. Build dependencies and test dependencies are typically used only during development and testing, and they do not affect the installation process. Therefore, in this paper, we only consider install dependencies and extra dependencies, which are the dependencies required for a successful installation of a package.
% Python uses \code{pip}~\cite{pip} as its official package manager for dependency management. Developers can specify all project dependencies in a file named \textit{requirement.txt} and use the command \code{pip install -r requirement.txt} to install them. In addition, developers can package and upload projects after declaring the dependencies in the configuration file. After compiling the configuration files, the metadata files are generated. The compiled dependency information is stored in the \textit{requires.txt} file of the source distribution package or in the \textit{METADATA} file of the binary distribution package. 

% Furthermore, Python supports various complex dependency types, including build, install, extra, and test dependencies. Each type can also be related to the local environment so that \pip will install them selectively. Build dependencies and test dependencies are typically used only during development and testing. They do not work in the download process. As a result, in this paper, we only consider install dependencies and extra dependencies since they are included in the dependencies required for a successful installation. 

\textbf{Module conflict examples.}
\begin{figure}[t]
    \centering
    \includegraphics[width=0.48\textwidth]{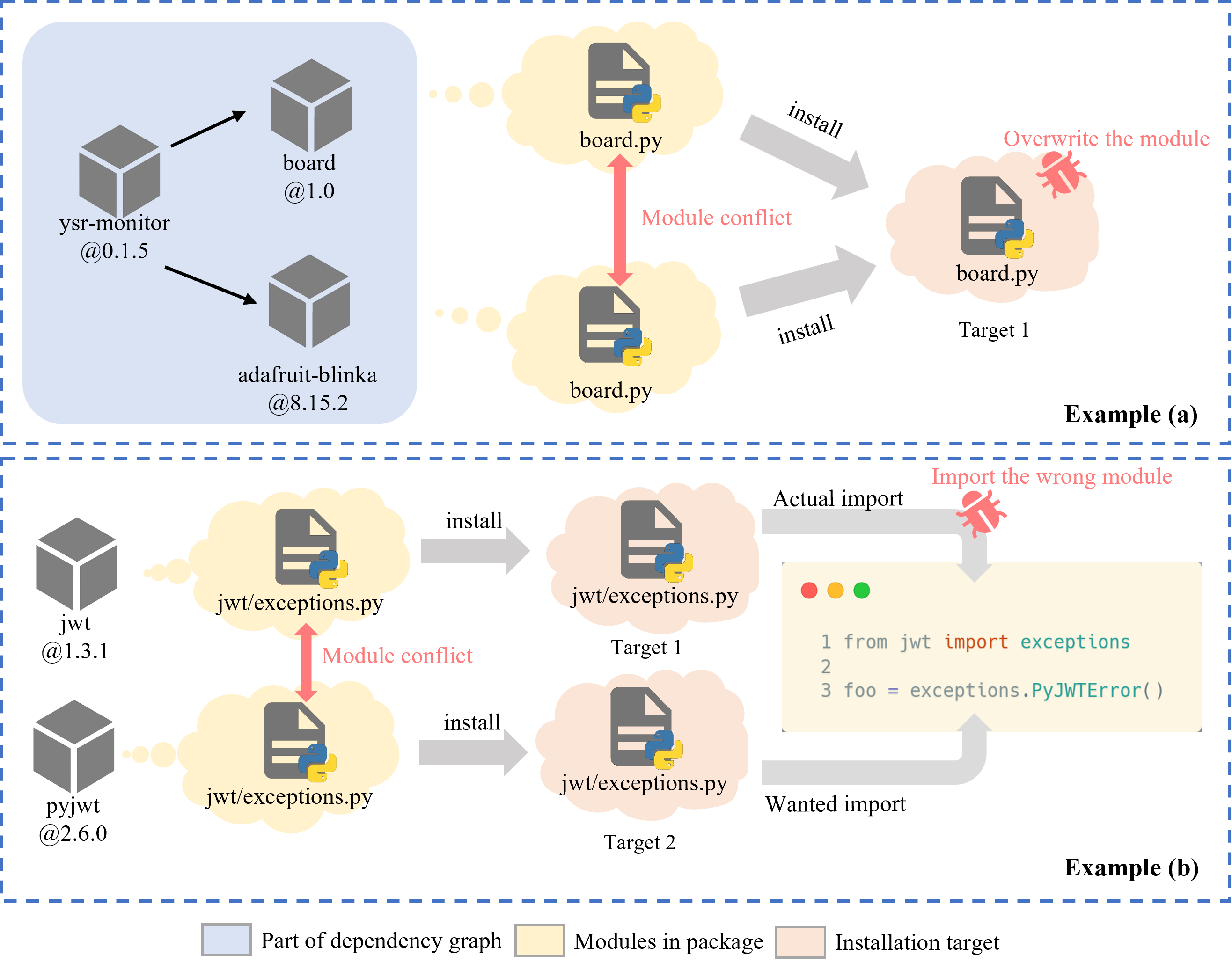}
    \caption{Module conflict example. Example (a) illustrates the overwriting module when downloading the package. Example (b) illustrates importing confusion when running the code.}
    \label{fig:motivating example}
\end{figure}
Here we give two examples of how module conflicts can affect Python projects.
\Cref{fig:motivating example}(a) shows a module conflict example that causes the module overwriting threat. More specifically, the project \code{ysr-monitor@0.1.5}~\cite{ysr-monitor} has two dependencies \code{board@1.0}~\cite{board} and \code{adafruit-blinka@8.15.2} ~\cite{Adafruit-Blinka}, which both contain a module named \code{board.py}. When \pip installs these packages, it will overwrite the \code{board.py} module from the first package with the one from the second package. This can lead to errors that are hard to debug, especially in CI/CD environments. To fix this issue, the developer needs to manually find the overwritten module and reinstall the corresponding package. However, this process is complicated and may corrupt the local environment further.
\Cref{fig:motivating example}(b) shows another example of importing a wrong module. The packages \code{jwt@1.3.1} and  \code{pyjwt@2.6.0} are installed in the two different targets (e.g., one in the system site-packages folder and one in the virtualenv~\cite{virtualenv} site-packages folder). However, they both have the \code{jwt/exceptions.py} module, which will conflict when imported. The Python interpreter will import the module based on the order of the folders in the \code{sys.path} list, which may not be the intended one. To fix this issue, the developer needs to either change the order of the folders in the \code{sys.path} list or use absolute paths to import the modules, which is inconvenient and error-prone.
\section{\name Design}
\label{sec:moduleguard}
\begin{figure*}[t]
    \centering
    \includegraphics[width=0.85\linewidth]{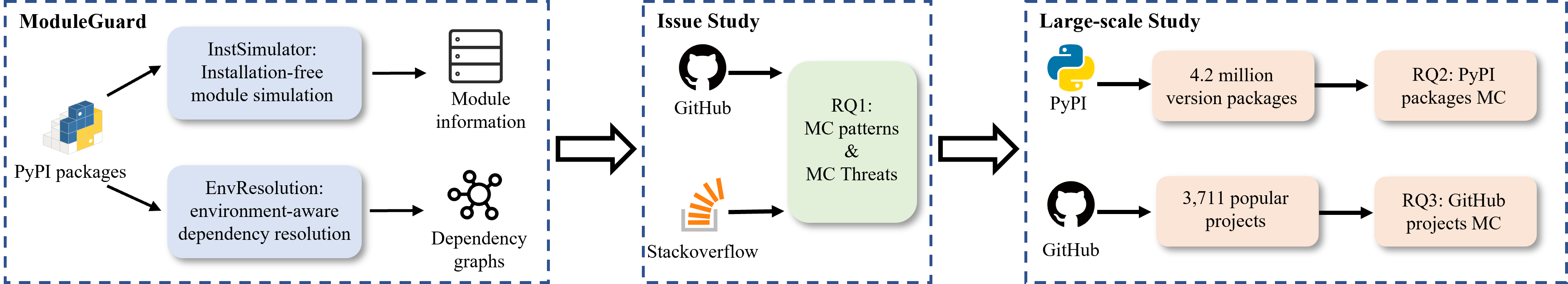}
    \vspace{5pt}
    \caption{Overview of our work.}
    \label{fig:workflow}
\end{figure*}

We present \name, a tool that helps us to investigate the impacts of module conflicts in the Python ecosystem. \Cref{fig:workflow} illustrates the framework of our work, which comprises two main components. The first component is \extractor, which extracts module information from various configuration files and simulates the installation process without actually installing the packages. The second component is \resolver, which resolves more accurate dependency graphs of Python packages by considering the local environment and the extra dependencies.

\subsection{\extractor: Installation-free Module Extraction}
\label{sec:InstExtractor}

\begin{figure}[t]
    \centering
    \includegraphics[width=\linewidth]{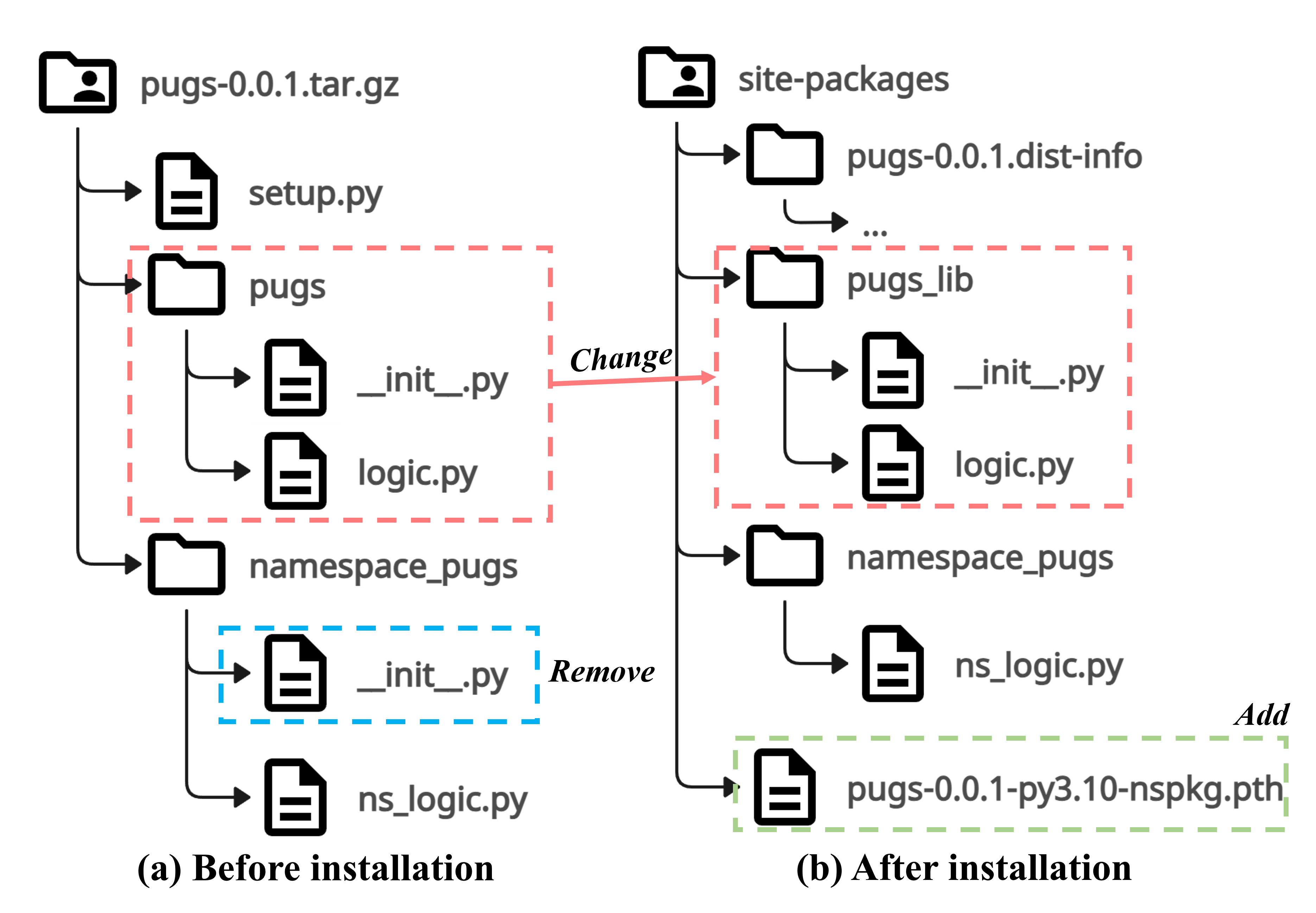}
    \caption{Module paths change after installation. Specific parameters in the configuration file control these behaviors.}
    \label{fig: Module paths change.}
\end{figure}

\textbf{Challenges.} To conduct a large-scale study of the Python ecosystem, we need to obtain the module information of Python packages. However, this is not a trivial task, as it faces several challenges. 
% The official package manager \pip acquires modules through decompression, building, and installation. However, this process is time-consuming and may fail due to local environment incompatibilities. Therefore, a static analysis tool that can parse module information efficiently and accurately is urgently needed. However, it faces several challenges. 
% 
First, PyPI has several types of packages, and each type has its own configuration files and formats. However, there is no comprehensive documentation that specifies which files and parameters are related to module information and how to parse them.
Second, the module information can change before and after the package installation. As shown in \cref{fig: Module paths change.}, the package \code{pugs@0.0.1} has different modules before and after the installation. This is because the installation process is controlled by configuration files (e.g., \code{setup.py}), which can add, remove, or rename modules. For example, if \code{namespace_packages = ['namespace_pugs']} is defined in the \code{setup.py}, it will remove the \code{namespace_pugs/__init__.py} module and add a new \code{nspkg.pth} module.
% Second, the module path can differ before and after the package is installed. As shown in \cref{fig: Module paths change.} , before and after the installation of \code{pugs@0.0.1}, the module has changed, removed,  and added behaviors. These behaviors are controlled by configuration files (i.e. setup.py). For example, if \code{namespace_packages = ['namespace_pugs']} is defined in the \code{setup.py}, it will remove the \code{namespace_pugs/__init__.py} module and add a new \code{nspkg.pth} module.
% For example, one of the module paths in is \code{src/pugs/logic.py} before installation and \code{pugs/logic.py} after installation. Moreover, some modules may be added or removed during installation (e.g. \code{tests/__init__.py}). 

% 
% To address these challenges, we propose \extractor, which has two main functionalities: (1) it parses the raw module data from different types of configuration and metadata files, and (2) it simulates the installation process to obtain the accurate module information without installing the packages.
To address these challenges, we propose \extractor, which has two main functionalities: (1) To solve the first challenge, we systematically study all module-related files and parse the raw module data from different types of configuration and metadata files, and (2) to solve the second challenge, we leverage a novel approach to simulate the installation process to obtain the accurate module information without installing the packages.

\begin{table}[t]
\renewcommand{\arraystretch}{1.1}
\centering
\caption{Module and dependency-related data.}
\label{tab:statistic data}
  \begin{threeparttable}
    \resizebox{0.95\columnwidth}{!}{
\begin{tabular}{cccc}
\hline
                                                    & File           & Module-Related Data                                                                                 & Dependency-Related Data                                                      \\ \hline
\multicolumn{1}{c|}{\multirow{3}{*}[-2ex]{\rotatebox{90}{\centering Metadata Files}}}          & EGG-INFO\tnote{*}       & \begin{tabular}[c]{@{}c@{}}top\_level.txt\\ SOURCES.txt\end{tabular}                                & requires.txt                                                                 \\ \cline{2-4} 
\multicolumn{1}{c|}{}                               & egg-info\tnote{*}        & \begin{tabular}[c]{@{}c@{}}top\_level.txt\\ namespace\_packages.txt\\ SOURCES.txt\end{tabular}      & requires.txt                                                                 \\ \cline{2-4} 
\multicolumn{1}{c|}{}                               & dist-info\tnote{*}       & \begin{tabular}[c]{@{}c@{}}top\_level.txt\\ namespace\_packages.txt\\ RECORD\end{tabular}           & METADATA                                                                     \\ \hline
\multicolumn{1}{c|}{\multirow{3}{*}[-2ex]{\rotatebox{90}{\centering Configuration Files}}} & setup.py       & \begin{tabular}[c]{@{}c@{}}py\_modules\\ packages\\ package\_dir\\ namespace\_packages\end{tabular} & \begin{tabular}[c]{@{}c@{}}install\_requires\\ extras\_require\end{tabular}  \\ \cline{2-4} 
\multicolumn{1}{c|}{}                               & setup.cfg      & \begin{tabular}[c]{@{}c@{}}py\_modules\\ packages\\ package\_dir\\ namespace\_packages\end{tabular} & \begin{tabular}[c]{@{}c@{}}install\_requires\\ extras\_require\end{tabular}  \\ \cline{2-4} 
\multicolumn{1}{c|}{}                               & pyproject.toml & \begin{tabular}[c]{@{}c@{}}py-modules\\ packages\\ package-dir\end{tabular}                         & \begin{tabular}[c]{@{}c@{}}dependencies\\ optional-dependencies\end{tabular} \\ \hline
\end{tabular}
}

\begin{tablenotes}
  \footnotesize
  \item[\hspace{0pt}*] The EGG-INFO is in "egg" type packages. The egg-info is in "tar.gz" \\ type packages. The dist-info is in "whl" type packages.
\end{tablenotes}
\end{threeparttable}
% \vspace{-5pt}
\end{table}

 % \newline   \code{packages=['pugs_lib', 'namespace_pugs'], package_dir={"pugs_lib": "pugs"}, namespace_packages = ['namespace_pugs'],}
\textbf{Raw module data extraction.} 
\label{chapt: Raw Module Data Parsing}
To extract the raw module data from Python packages, we first identify the types and formats of files that contain module information. Then, we implement different parsers for each type of file. 

We conduct a systematic study of the packaging, compilation, and installation process of Python packages. We use differential testing to modify different parameters in the configuration files and observe the changes in the metadata files and the module information. Based on these results, we combine the official Python~\cite{pydocumentation} and pip~\cite{pip} documentation to infer the function of different parameters.
\Cref{tab:statistic data} shows the result of our study. Python packages have three kinds of metadata files: \code{egg-info}, \code{EGG-INFO}, and \code{dist-info}. They correspond to the \code{.tar.gz}, \code{.egg}, and \code{.whl} distribution packages, respectively.
Moreover, Python packages have three types of configuration files: \code{setup.py}, \code{setup.cfg}, and \code{pyproject.toml}. The \code{setup.py} is an executable script file, while the other two are formatted configuration files. They can all define module information but use different formats and parameters.
% We have implemented a simple demo to understand a Python package's packaging, compilation, and installation process. Then we use differential testing to modify different parameters one by one and observe the generated metadata files and the module information after installation. Based on these results, we combined the Python\cite{pydocumentation} and pip\cite{pip} documentation, to infer the function of different parameters.
% \Cref{tab:statistic data} lists the result of our experiments. Python packages have three kinds of metadata files: \code{egg-info}, \code{EGG-INFO}, and \code{dist-info}. They correspond to the \code{.tar.gz}, \code{.egg}, and \code{.whl} distribution packages, respectively.
% %
% What's more, Python packages have three types of configuration files: \code{setup.py}, \code{setup.cfg}, and \code{pyproject.toml}. The \code{setup.py} is an executable script file, while the other two are formatted configuration files. They can all define module information but use different formats and parameters.
% 

The \extractor implements different parsers for each type of configuration and metadata file. Specifically, \extractor converts text-type files into a list, parses formatted configuration files with a format-specific parser, and for \code{setup.py}, the executable script file, \extractor uses AST and data flow analysis to extract module configuration parameters based on a PyCD tool proposed in~\cite{cao2022towards}. These parsers can extract the relevant information from the files and store them in a structured way.
In this way, \extractor can avoid decompressing the package locally by reading only a few configuration files and metadata files from memory, which saves time and space.

\textbf{Installation-free module simulation.}
We propose a novel technique named \textit{installation-free module simulation} to convert raw module data into module information after installation. This technique consists of three steps:
%The root node of the tree represents the target folder to which the package needs to be installed, the middle nodes represent the folders in the package, and the leaf nodes represent the modules in the package. 
(1) \extractor first takes the file structure in the compressed package and turns it into a virtual file tree. (2) It then translates each raw data semantics into operations that add, delete, and search for nodes in the virtual file tree. 
(3) It employs the DFS algorithm to traverse the file tree and obtain all module paths.

For example, suppose a package has the raw data \code{packages = [pugs_lib, namespace_pugs]} and \code{package_dir = pugs_lib : pugs}. The \code{packages} data is a list of folder names after installation, and the \code{package_dir} data is a mapping of folder names before and after installation (\code{pugs_lib} is the folder name of the \code{pugs} after installation). If there is no mapping in \code{package_dir}, the folder name will not change before and after installation.
Based on their semantics, \extractor uses the BFS algorithm to search for the folder names (e.g., \code{pugs}) before installation recorded in \code{package_dir} (searching root node is the parent node of the \code{setup.py} or other configuration files), and changes their node names to the names after installation (i.e., \code{pugs_lib}). Then, it deletes all the subtrees except the nodes recorded in the \code{packages} data. After simulating all the raw module data semantics, we can obtain a file tree that contains the module information without the installation process. Finally, \extractor uses the DFS algorithm to traverse the file tree and obtain the module paths. Note that each path from the root node to a leaf node within this tree corresponds to a module path. We have described the semantics of all the parameters and a more detailed code demo in the artifact we release.

\subsection{\resolver: Environment-aware Dependency Resolution}
\label{sec:DGR}
\textbf{Challenges.} To investigate the module conflicts in the Python ecosystem, we need to resolve the dependency graphs of over 4.2 million packages. However, it faces the following challenges.
% Since \pip installs dependencies during the installation process, these dependencies also may have module conflicts. So we need to resolve the dependency graph for over 4.2 million packages. However, it faces the following challenges. 
%
% First, the Python package index has structural limitations that prevent \pip from obtaining dependency information without downloading and extracting packages, which slows down its dependency resolution. Thus parsing the entire ecosystem packages has an intolerable time overhead. 
%
First, using common static resolution methods to obtain the dependency graphs without installation is not very accurate. This is because the Python dependency graphs depend on the local environment information, and the dependencies between packages are complex due to the extra dependencies.
Moreover, previous work~\cite{wang2020watchman, depsdev, Wang2022SmartPip, Libio} either lacks up-to-date information or has lower accuracy. They often parse dependencies from only a single dimension, such as \code{setup.py} or \code{requires.txt}, resulting in incomplete information.
Furthermore, they do not consider the local environment-related and extra dependencies, which are common in Python packages and can affect the dependency graphs. This leads to inaccurate dependency graphs. On the other hand, obtaining accurate dependency graphs using \pip requires the installation process, which is time-consuming and may fail due to local environment incompatibilities.

To solve these, we present \resolver, a local environment-aware dependency resolution that supports the local environment and extra dependencies. It consists of three steps: (1) multidimensional dependency information extraction, (2) local environment information collection, and (3) dependency graph resolution.

\textbf{Multidimensional dependency information extraction.} \resolver adopts a multi-dimensional approach for extracting direct dependencies from three dimensions: PyPI API~\cite{pypi-api}, dependencies in metadata files, and dependencies in configuration files. Similar to the approach in \cref{chapt: Raw Module Data Parsing}, we first classify the dependency-related information from the metadata and configuration files, as shown in \cref{tab:statistic data}. In this step, we will save both the direct dependencies with their environmental conditions and the optional extra dependencies of the Python project. Then we convert the dependency information parsed from different files into a unified format for subsequent parsing.
% Then we extract the dependency information and unify them into a triplet format $D = (N_d, C_d, M)$, where $N_d$ is the dependency name, $C_d$ is its constraint, and $M$ is environmental information. 
% % The value of $M$ can be null, otherwise, it will be a pair: $M = (N_m, C_m)$, where $N_m$ denotes the local environmental variable name and $C_m$ denotes environmental constraint. For example, $D1 = (numpy, >= 1.21.0, (python\_version, >= 3.11))$ represents that \code{pandas@1.5.3} has a dependency \code{numpy} if the condition \code{python_version >= 3.11} is satisfied. 
% Moreover, we denote extra dependency as a special local environment condition since their rules are similar. For example, $D2 = (brotlipy, >= 0.7.0, (extra, == 'compression'))$ represents that \code{pandas@1.5.3} would has a dependency \code{brotlipy} if another package declares a \code{pandas[compression]} dependency.

\textbf{Local environment information collection.} For the local environment, \resolver collects 11 types of environmental information~\cite{depspecifier} that may affect the dependency graphs, such as \code{python_version}, \code{os_name}, and so on. These environment variables and their values are stored in a global dictionary when resolving dependencies (e.g., \code{{(python_version, 3.10), (os_name, posix), (sys_platform, linux), ...}}).

\textbf{Dependency resolution.} \resolver uses the resolvelib~\cite{resolvelib} framework as the core backtracking resolution algorithm. To improve the efficiency and accuracy of dependency graph resolution, \resolver implements the following optimizations. 
First, like previous work~\cite{wang2020watchman, cao2022towards, Wang2022SmartPip}, \resolver also employs a local knowledge base. However, \resolver adopts multi-dimensional dependency information extraction, which can obtain more comprehensive dependency information than previous work, including local environmental conditions and extra dependencies. 

Second, \resolver supports resolving extra dependencies and local environment dependencies. More specifically, during dependency resolution, an extra dependency will add an entry to the environment variable dictionary. For example, dependency \code{pandas[compression]} will add \code{compression} to the environmental variable dictionary. After that, we can treat an extra dependency as a special local environmental dependency. When resolving each direct dependency of packages, \resolver checks whether the dependency's environmental conditions match the values in the environmental variable dictionary. A dependency will be dropped if the value in the dictionary does not meet the condition. For instance, dependency \code{numpy>= 1.21.0; (python_version, >= 3.11)} will be dropped if the python version is \code{3.10} and does not meet the condition \code{>=3.11}.

Third, \resolver adopts a priority policy when resolving dependencies for efficiency. This is based on our observation that the order of dependencies does not affect the result of the resolution, but it does affect the time to backtrack the resolution algorithm. A good order can reduce the number of backtrack times, thus improving the parsing efficiency. Therefore, \resolver sorts the dependencies according to the following rules in each recursion: resolving the pinned version dependencies first, then the dependencies with a scope constraint, and finally the dependencies with no constraint. Moreover, dependencies that are close to the root node are always resolved before dependencies that are far from the root node. In this way, \resolver can significantly reduce the number of backtracking times and thus improve the resolution efficiency.

% What's more, \resolver adopts a priority policy when resolving dependencies for efficiency. This is mainly based on our observation that the order of dependencies does not affect the result of the resolution, but it does affect the time to backtrack the resolution algorithm. A good order can reduce the number of backtrack times, thus improving the parsing efficiency. As a result, \resolver sorts the dependencies according to the following rules in each recursion: resolving the pinned version dependencies first, then the dependencies with a scope constraint, and finally the dependencies with no constraint. What's more, dependencies that are close to the root node are always resolved before dependencies that are far from the root node.

\begin{table}[t]
\renewcommand{\arraystretch}{1.1}

\centering
\caption{Evaluation of \name.}
\label{tab:evaluation}
\begin{threeparttable}
     \resizebox{0.98\columnwidth}{!}{
\begin{tabular}{ccccccc}
\hline
Benchmark                                                                                                                    & Dataset & Correct & Miss & Excess & Error & Accuracy \\ \hline
\multicolumn{1}{c|}{\multirow{2}{*}{\begin{tabular}[c]{@{}c@{}}\ \extractor\end{tabular}}}         & Data1   & 4,045    & 152     & 28   & 7     & 95.58\%  \\
\multicolumn{1}{c|}{}                                                                                                        & Data2  & 3,834    & 116     & 37   & 2    & 96.11\%  \\ \hline
\multicolumn{1}{c|}{\multirow{2}{*}{\begin{tabular}[c]{@{}c@{}} \resolver\\ (Node)\end{tabular}}} & Data1   & 4,177    & 41     & 8    & 13    & 98.70\%  \\
\multicolumn{1}{c|}{}                                                                                                        & Data2  & 3,795    & 93     & 30    & 20    & 96.37\%  \\ \hline
\multicolumn{1}{c|}{\multirow{2}{*}{\begin{tabular}[c]{@{}c@{}} \resolver\\ (Edge)\end{tabular}}} & Data1   & 4,133    & 46     & 11   & 47    & 97.66\%  \\
\multicolumn{1}{c|}{}                                                                                                        & Data2  & 3,748    & 107     & 40    & 33    & 95.18\%  \\ \hline
\end{tabular}
}
\begin{tablenotes}
    \item[\hspace{0pt}*] Node: evaluate nodes in dependency graph only. 
    \item[\hspace{0pt}*] Edge: evaluate nodes and edges in the dependency graph. 
    \item[\hspace{0pt}*] Accuracy: correct/total
\end{tablenotes}
\end{threeparttable}
\end{table}

\section{\name Evaluation}
\label{sec:evaluation}
% We propose the following research question for evaluating the \name.
% \begin{itemize}
% \item \textbf{RQ1(accuracy)} How accurate is \name in module simulation and dependency resolution?
% \end{itemize}
Due to the lack of established ground truth for comparison, we carefully construct a benchmark to evaluate the two aspects of \name, respectively. 
We collect datasets from the two following sources:
\begin{itemize}[leftmargin=*]
    \item Dataset 1. We select the top 3,000 projects from Libraries.io~\cite{libquery} and PyPI Downloads Table~\cite{PyPI-table} for six months from August 2022 to February 2023, respectively and we apply a de-weighting process to the two sets to reduce the bias. 
    \item Dataset 2. We randomly select 5000 projects from the total list of the PyPI package.
\end{itemize}
% 1605 projects declare extra dependency， 382 projects use extra dependency, 3515 projects have dependency%
For each dataset, we select the latest version and install it using \pip for each project. We use the module relative paths and the dependency graphs obtained after the installation as the ground truth for comparison. There are two main reasons for installation failure: First, the local environment is not compatible with the package, e.g. Python2 package cannot be installed in the Python3 environment; Second, an error occurred while running the installation script, causing the installation process to exit. Finally, we get 4,232 and 3,989 projects in the two datasets.

\textbf{Evaluate metrics.}
We define four metrics to evaluate the accuracy of \name. (1) Correct. The modules or dependency graphs resolved by \name are totally consistent with the ground truth. (2) Miss. Some modules or some elements in dependency graphs of the ground truth do not exist in our results. (3) Excess. Some modules or some elements in dependency graphs resolved by \name do not exist in the ground truth. (4) Error. Other cases. 

\textbf{Experimental setup.} To obtain the module paths after installation, we use \code{pip install XXX -t target --no-dependencies} to install packages. This command implies that the latest version package will be installed in the target folder and no dependencies will be installed. Moreover, we only considered modules with \code{.py} extensions, so the data files (e.g. pictures, tables) included in the packages will be ignored. 

To obtain the dependency graphs, we also use \pip installation to get the ground truth. We add the following settings to obtain the exact dependency graph. First, \pip installation process depends on the repository status of the remote, which is updated in real-time. In contrast, our local knowledge base is updated on a daily basis. To address this gap, we mirrored approximately 13TB of PyPI packages locally with bandersnatch~\cite{bandersnatch} tool and we use this local mirror during \pip installations (i.e. \code{pip install -i localhost/simple}). Second, in order to obtain \pip's dependency resolution results, we hook \pip's dependency resolve function and write the dependency graph to files. 
% we get 4,232 and 3989 projects in the two datasets.
% Finally, we successfully installed 2784 and 2828 packages for each benchmark, respectively. There are two main reasons for installation failure: First, the local environment is not compatible with the package, e.g. Python2 package cannot be installed in the Python3 environment; Second, an error occurred while running the installation script, causing the installation process to exit.

\textbf{\extractor result.} \Cref{tab:evaluation} shows the results of \extractor evaluation benchmark. The results show that \extractor has the ability to extract module information with \textbf{95.58\%} and \textbf{96.11\%} accuracy on different datasets. 
In addition, the table shows that \extractor technique has 152 (3.59\%) and 116 (2.91\%) Misses on the two datasets, respectively. This is mainly because \extractor uses AST static analysis method parsing the \code{setup.py} install script, which has limitations on parsing syntactically complex install scripts correctly. 
Moreover, a given version of a Python project has multiple packages with slightly different modules. However, the \extractor selects only one of these packages to parse, which might differ from the package actually installed. For instance, the \code{0.4.4} version of the \code{jaxlib} project has 12 packages.
Overall, \name is able to achieve over 95\% accuracy in module extraction. It meets our requirement of extracting the whole ecosystem packages' modules.

\textbf{\resolver result.} 
We considered both node-level and graph-level benchmarks to evaluate the accuracy of the \resolver technique in \name. The node-level benchmark focuses solely on the individual nodes within the graph, without considering the edges between them. It evaluates the ability to extract direct dependencies. In contrast, the graph-level benchmark takes into account both the nodes and edges present in the graph. It evaluates the ability to resolve dependency graphs. 

\Cref{tab:evaluation} shows the results of our evaluation. 
The results show that the accuracy of \resolver ranges from 95.18\% to 98.70\% where the edge accuracy is lower than that of node.
What's more, we manually reviewed projects that were unable to resolve and identified two primary causes for their failure. First, we are unable to process complex \code{setup.py} configuration files. For example, the \code{setup.py} file of the project \code{ta} retrieves dependencies from files within the package, however, \name does not unpack the package to optimize extraction speed. Second, we read dependencies from metadata files generated by maintainers' local environments, which may deviate from our experimental environment. For instance, project \code{fortnitepy} has three direct dependencies in its metadata, whereas pip only extracts two during our local installation. 

Overall, the results demonstrate that although \resolver has some limitations in extracting dependencies information, it is capable of handling large-scale packages quickly with an acceptable error.
\section{Large-scale Study}
\label{sec:empirical-study}
Since the MC problem has not been studied systematically in Python language in existing work, we empirically study the MC issues from GitHub and Stack Overflow 
 and classify them into three patterns. We then used \name to evaluate all 4.2 million PyPI packages and 3,711 high-star projects collected from GitHub for the presence of MCs and their potential impacts. In summary, we propose the following research questions:
\begin{itemize}[leftmargin=*]
\item \textbf{RQ1 (Issue Study).} What are the common types of module conflict issues? What potential threats might they have?
\item \textbf{RQ2 (PyPI Packages).} How many of all PyPI packages have MC effects?
\item \textbf{RQ3 (GitHub Projects).} How many popular projects on GitHub are affected by MC, and what are their characteristics?
\end{itemize}

\subsection{RQ1: Issue Study}
\subsubsection{Data Collection.}
We collect 97 MC issues in total and we search them in two steps. 
% How we collect
First, we combined two sets of keywords---(module OR name) AND (clash OR conflict) to search for MC issues on GitHub and added \code{is:issue} and \code{language:python} options.
Since Github can only show the first 100 pages of search results for each combination search result, we obtained the search results in order of best match and collected 4,000 issues. 
Second, for the 4,000 issues, the three co-authors manually reviewed the descriptions and bug reports in the issues and finally filtered out 55 issues that were strongly related to MC issues. We also notice that some maintainers or reporters would cite related issues in their comments. As a result, we searched for other issues mentioned in these 55 issues using the snowballing technique~\cite{ladisa2022taxonomy} and checked them manually. Finally, we collected 78 MC issues from GitHub.
The keyword "Python module name (clash OR conflict)" was used to search on StackOverflow. We manually review the top 200 most relevant issues that include answers. Ultimately, a total of 19 issues related to MC are collected from Stackoverflow.

\subsubsection{Module Conflict Types.}
After studying the collected 97 MC issues, we observe that module conflicts can occur in three situations after packaging the project and uploading it to the PyPI or Github. First, modules of the project may conflict with the built-in standard library modules, causing \textit{module-to-Lib} conflict. Second, As a TPL, its modules can conflict with the other TPLs (not relevant to this project), leading to \textit{module-to-TPL} conflict. In addition, projects that declare direct dependencies in their configuration files may have module conflicts within the dependency graph (those TPLs are relevant), resulting in \textit{module-in-Dep} conflicts.

In the following, we give each type of conflict a formal definition and discuss them in detail with illustrative issue examples. 
To ease the discussion, we first give some grammar below:
\begin{displaymath}
\begin{aligned}
P &: \text{the set of all packages on PyPI}\\
p &: p\in P, \text{representing a specific package}\\
M_p &: \{m \mid m \text{ is a module after installing package } p\} \\
M_{Lib} &: \{m_l \mid m_l \text{ is a standard library module}\} \\
DG(p) &: \text{the dependency graph of the package}\ p \\
m_1 \xleftrightarrow{\text{conflict}} m_2 &: m_1 \text{ and } m_2 \text{ have the same name or same path}
\end{aligned}
\end{displaymath}

\textbf{Module-to-Lib conflict.} Conflicts can occur between the project's modules and standard library (Lib) modules (21/97=21.65\%). Suppose there are two modules. One module $m$ belongs to package $p$ and the other is a library module $M_{Lib}$ and they have conflicts. In that case, we consider the packages $p$ to have a module-to-Lib conflict. 
We formulate it as the following:
\begin{displaymath}
\begin{aligned}
\exists \ p \in P, & \\ m &\in M_p \land m_l \in M_{Lib} \land m \xleftrightarrow{\text{conflict}} m_l 
\end{aligned}
\vspace{-5pt}
\end{displaymath}

For example, the \#14 issue~\cite{14} of the \code{python-hgijson}. The package \code{python-hgijson@1.5.0} has a \code{json} module, and it conflicts with the standard library \code{json} module, resulting in module-to-Lib conflict. 
\vspace{+10pt}

\textbf{Module-to-TPL conflict.} Modules will also conflict with the modules of the other unrelated third-party packages (64/97=65.98\%). Suppose there are two different modules. One module $m$ belongs to package $p$ and the other module $m'$ belongs to another unrelated package $p'$ such that the two module conflicts with each other, we consider the package $p$ and $p'$ both have a module-to-TPL conflict. 
We formulate module-to-TPL conflict as the following:
\begin{displaymath}
\begin{aligned}
\exists \ p, p' \in & P \land p \neq p', \\ m &\in M_p \land m' \in M_{p'} \land m \xleftrightarrow{\text{conflict}} m'
\end{aligned}
\vspace{-5pt}
\end{displaymath}

For example, the \#3 issue~\cite{3} shows that the \code{python-slugify@8.0.0} and the \code{awesome-slugify@1.6.5} packages both have a \code{slugify} conflicting module and they have a conflict with each other if they installed together, so the two packages have a module-to-TPL conflict.

\vspace{+10pt}

\textbf{Module-in-Dep conflict.} Conflicts can occur within the project's dependency graphs (12/97=12.37\%). If there are two packages $p$ and $p'$ within a Dependency Graph of the root package $r$ ($r$ can be one of $p$ or $p'$, or it can be another package), a module $m$ within package $p$ and a module $m'$ within package $p'$ such that they conflict with each other, we denote the packages $r$ has a module-in-Dep conflict. 
We formulate it as the following:
\begin{displaymath}
\begin{aligned}
\exists \ (p, p') \subseteq & DG(r) \land p\neq p', \\ m & \in M_p \land m' \in M_{p'} \land m \xleftrightarrow{\text{conflict}} m'
\end{aligned}
\vspace{-5pt}
\end{displaymath}

For example, the \#44 issue~\cite{44} of the\code{emoca}. The package \code{emoca@1.0} has \code{opencv-python-headless@4.5.5} and \code{opencv-python@4.5.5} in its dependency graph, and they both have a \code{cv2} module, so the root package \code{emoca@1.0} has a module-in-Dep conflict.  

\begin{table}[t] 
\renewcommand{\arraystretch}{1.2}
\centering
\caption{Statistics of MC types and their potential threats.}
\label{tab:statistic issues}
\begin{threeparttable}
    \resizebox{0.85\columnwidth}{!}{
\begin{tabular}{ccc}
\hline
MC types    & Modules overwriting & Import confusion \\ \hline
Module-to-Lib (21)      & -        & \checkmark                 \\ 
Module-to-TPL (64)      & \checkmark        & \checkmark                 \\
Module-in-Dep (12)       & \checkmark        & -                \\ \hline
% Summary (97)                    & 63       & 2               \\ \hline
\end{tabular}
}
\end{threeparttable}
\end{table}

\subsubsection{Module Conflict Threats}
Based on the description of issues, we summarize that the Python ecosystem suffers from two shortcomings in code management. First, \pip installs TPL modules into the \code{site-packages} folder by default and does not isolate packages from each other. This means that different packages, including their direct and indirect dependencies, are mixed in the same directory. As a result, different modules from different TPLs would conflict with each other and will cause \textbf{modules overwriting} threats. Second, Python provides a more flexible code management mechanism, allowing the import of modules (with access rights) from anywhere in the system, including standard libraries, TPL modules, and the project's own modules. However, they use the same statement for importing and do not differentiate between them. The Python interpreter searches for modules using the first-match principle. Therefore, conflicts can occur between modules that have the same module path but are installed in different locations and will lead to \textbf{importing confusion} threats. 
Furthermore, these two flaws pose threats that will have potential impacts on the installation, upgrade, and importing of software packages. \Cref{tab:statistic issues} illustrates statistics on the different types of module conflicts that may have threats on the corresponding stages. In the following, we will introduce them in detail with examples. 

\textbf{Threat 1 (Modules overwriting).} 
Module overwriting is a serious threat to the integrity and functionality of Python projects, as it may cause unexpected errors or behaviors. Module overwriting occurs when two modules with the same relative module path are installed into the same directory, resulting in one module being overwritten by another. It can be triggered by two types of module conflicts: module-to-TPL conflict and module-in-Dep conflict.

Module-to-TPL conflict refers to the situation where two packages have conflicting modules. This can happen when \pip installs TPL modules into the \code{site-packages} folder by default and it does not isolate packages from each other. For instance, in issue \#4625~\cite{4625} of the project \code{pypa/pip}, the developer installs both \code{pyjwt@1.5} and \code{jwt@0.5.2}, which have conflicts on the module \code{jwt/exceptions.py}. The later installed module will overwrite the first installed module when two conflicting modules are installed simultaneously.

Moreover, module-to-TPL conflict can also occur in Windows systems due to case insensitivity of paths. For example, in issue \#156~\cite{156} of the project \code{pycrypto}, the developer installs the \code{crypto} (lowercase) module first, and then installs \code{pycrypto}, which has the \code{Crypto} (upper case) module name. However, since the \code{crypto} folder already exists locally, \pip cannot create the \code{Crypto} (upper case) folder and instead installs all modules under the pre-existing \code{crypto} (lowercase) folder. Consequently, this process overwrites \code{crypto}'s modules and breaks the project's functionality.

Module-in-Dep conflict refers to the situation where a project and its dependencies have conflicting modules. The conflict can cause module overwriting during the installation. For example, in issue \#841\cite{841} of the project \code{Albumentations}, the project installs both \code{opencv-python} (indirect dependency) and \code{opencv-python-headless} (direct dependency). They have conflicts on the \code{cv2} module and its sub-modules. Although the official documentation~\cite{opencv} states that they cannot be installed simultaneously, developers are not aware of this restriction as indirect dependencies are a black box for them.

Furthermore, module overwriting can also affect package upgrades, as \pip's update process may cause module conflicts during installation or uninstallation. For example, in issue \#8509~\cite{8509}, the package \code{ansible@2.9.10} misses some files after upgrading due to modules being overwritten during the update process. The update process is as follows: first, \pip installs the package's dependencies (where module overwrites may occur), then it uninstalls the old version of the package (possibly uninstalling newly installed modules that have already been overwritten), and finally it installs the new version of the package (package integrity is compromised).

% For the above threat to have any real impact, it requires that the conflicting modules of module-to-TPL be installed at the same time. On the other hand, the module-in-Dep conflict also increases the likelihood that the threat will have a real impact since both the project and the dependencies are installed at the same time in a single installation.

\textbf{Threat 2 (Importing confusion).} 
Conflicts between modules and the standard library (module-to-Lib) or third-party libraries (module-to-TPL) pose a potential threat of importing confusion. For module-to-Lib conflicts, the standard library modules are stored separately and cannot be overwritten when downloading a package. However, if a module has the same name as a standard library module, it can confuse the interpreter and cause it to import the wrong module. For example, the project \code{FibexConverter}~\cite{7} has a module named \code{parser.py}, which conflicts with the standard library module \code{parser}. This leads to an issue where the program imports the \code{parser} module from the standard library instead of the \code{parser.py} module from the project. This is because Python's import mechanism first searches for modules already imported in \code{sys.modules}, which contains a cache of modules pre-recorded when the Python interpreter starts up, such as os, abc, etc. As a result, modules with names identical to these standard library modules are not properly imported.

For module-to-TPL conflicts, Python searches for modules in the order of the paths in \code{sys.path} and stops at the first match it finds. When conflicting modules are downloaded and located in different locations, this requires the developer to be very experienced and carefully set the order of \code{sys.path} to handle conflicts. Note that while the namespace package handles modules overwriting, it does not address the importing confusion threats caused by the import prioritization. Moreover, Python also supports various ways to install packages, which often have different default paths. For example, one can use apt-get to install packages in \code{/usr/}, \pip to install packages in \code{site-packages/}, or other tools such as conda~\cite{conda} and poetry~\cite{poetry} that have their own default installation paths. This greatly increases the threat of importing confusion.

\subsection{RQ2: PyPI Packages Study}
\label{sec:large-scale-study}
We use \name to conduct a large-scale study of module conflicts (MCs) in the PyPI ecosystem. We study three types of MC patterns for all 4.2 million PyPI packages as of March 2023.
For module-to-TPL conflict, we only consider the latest version packages for each project as of March 2023, since different versions of the same project cannot coexist in a local environment and \pip will install the latest version by default unless specified constraints otherwise. We identify packages that have module conflicts, and make the assumption that these packages will be installed at the same time.
For module-to-Lib conflict, we first collect 199 standard library module names from the Python official documentation~\cite{pylibdocumentation}. Then we analyze the module names used by all the packages in the ecosystem. It's worth noting that we cannot know the order of \code{sys.path} or the standard library in the users' environment. Therefore, we also assume that the users have 199 standard libraries available locally, all of which are loaded into the cache.
% We note that using standard library names does not necessarily imply a threat of import confusion in practice, since it depends on the availability and cache of the standard libraries when the user runs the program. However, we conservatively assume that these packages have potential impacts.
For module-in-Dep conflict, we consider all version packages for each project and resolve their dependency graphs with \resolver. For the nodes in the resolved dependency graphs, we check whether their modules have conflicts. 

We extract 177,216,363 modules and 27,678,668 direct dependencies for 4,223,950 packages from PyPI as of March 2023 and resolve 4,223,950 dependency graphs. This includes 424,823 latest version packages with 5,419,306 modules.

\begin{table}[t]
\small
\centering
\renewcommand{\arraystretch}{1.1}
\caption{Top 10 conflict modules in packages.}
\label{tab:Top 10 module paths}
\begin{threeparttable}
    \resizebox{0.95\columnwidth}{!}{%
    \begin{tabular}{lcc}
\hline
\multicolumn{1}{c}{Module paths}      & \# of latest pkgs  & \# of all pkgs\\ \hline
src/\_\_init\_\_.py                   & 1,157               & 8,777\\
\_\_init\_\_.py                       & 1,083               & 4,421\\
utils/\_\_init\_\_.py                 & 410                & 3,899\\
distributions/\_\_init\_\_.py         & 404                & 448\\
distributions/Generaldistribution.py  & 394                & 431\\
distributions/Gaussiandistribution.py & 394                & 431\\
distributions/Binomialdistribution.py & 393                & 428\\
client/\_\_init\_\_.py                & 367                & 1,142\\
scripts/\_\_init\_\_.py               & 363                & 5,336\\ 
server/\_\_init\_\_.py               & 360                & 796\\ \hline
\end{tabular}
}
\end{threeparttable}
\end{table}

\textbf{Module-to-TPL conflict.} 
We use the latest version packages of 424,408 projects as of 2023 March to study module-to-TPL conflicts. We find that 91,134 (21.45\%) packages have module-to-TPL conflicts, affecting 386,595 (7.13\% out of 5,419,306) module paths. These packages may have \textit{module overwriting} or \textit{importing confusion} threats depending on whether they are installed in the same or different locations. Moreover, 27,851 (6.56\%) packages may have an overwriting impact in a Windows environment, involving 3,517 module paths.

\textbf{Findings.} 
We observe that developers often package redundant modules that are not needed for runtime, such as testing modules (e.g., 41,095 packages have \code{test(s)/__init__.py},) and example modules (e.g., 14,877 packages have \code{example(s)/__init__.py}). These modules are only for the development process and are more error-prone and confused~\cite{docsertuparg}. They not only increase the storage pressure on the PyPI server, but also slow down the efficiency of \pip resolution due to the backtracking algorithm.

Furthermore, we identify the top 10 most common module paths in software packages as shown in \cref{tab:Top 10 module paths}. There are over 1000 packages that include \code{src/__init__.py} and \code{__init__.py}, which are the result of the misconfiguration of the src-layout and flat-layout format packages~\cite{docsertuparg}, respectively. These two modules are stored in the project root directory without any meaning or functionality. 

Additionally, we find that packages with conflicting modules often have similar names, which reflect their functionality. For example, out of the 404 packages that have the \code{distributions/__init__.py} module, 290 contain the substring 'distribution' in their project name. This means that conflicting packages are more likely to be installed together, because they most likely belong to the same domain or have the same functionality.

In addition, through exploring related GitHub Issues, we find that project maintainers who have the same module name are often reluctant to change their own module name. Changing the module name will not only break forward compatibility, but also the workload is very large, and increase the learning cost of users when used.

% \begin{table}
% \centering
% [32966, 23331, 19447, 14452, 9464, 7020, 5227, 2888, 2145, 2567, 1912, 799, 744, 258, 77, 14, 25, 1]
% \begin{tabular}{l l l l} 
% 2023 & 191659 & 15435 & 32966 & 0.080534 \\
% 2022 & 957388 & 87362 & 23331 & 0.09125 \\
% 2021 & 853055 & 71063 & 19447 & 0.083304 \\
% 2020 & 676009 & 57997 & 14452 & 0.085793 \\
% 2019 & 421086 & 33126 & 9464 & 0.078668 \\
% 2018 & 304270 & 25116 & 7020 & 0.082545 \\
% 2017 & 236270 & 18215 & 5227 & 0.077094 \\
% 2016 & 168821 & 13951 & 2888 & 0.082638 \\
% 2015 & 125265 & 8945 & 2145 & 0.071409 \\
% 2014 & 88486 & 5829 & 2567 & 0.065875 \\
% 2013 & 62461 & 3416 & 1912 & 0.05469 \\
% 2012 & 39426 & 1934 & 799 & 0.049054 \\
% 2011 & 25075 & 1241 & 744 & 0.049492 \\
% 2010 & 17674 & 624 & 258 & 0.035306 \\
% 2009 & 12259 & 476 & 77 & 0.038829 \\
% 2008 & 6641 & 228 & 14 & 0.034332 \\
% 2007 & 2922 & 74 & 25 & 0.025325 \\
% 2006 & 1093 & 35 & 1 & 0.032022 \\
% 2005 & 240 & 1 &  & 0.004167 \\
% 1970 & 33850 &  &  \\
% \end{tabular}
% \end{table}
\begin{figure}[t]
    \centering
    \includegraphics[width=0.85\linewidth]{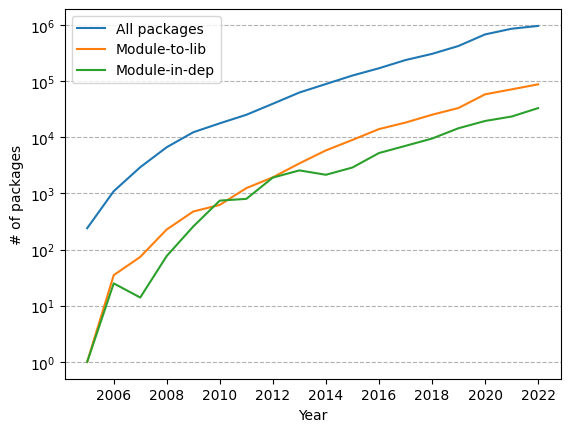}
    \caption{Statistics of the number of packages released and the number of conflict packages in each year.}
    \label{fig: Libtimestamp}
\end{figure}
\textbf{Module-to-Lib conflict.}
We analyzed the entire ecosystem of 4.2 million packages and found that 345,068 (8.17\%) packages have module-to-Lib conflicts, which may cause import errors at runtime. Moreover, we discovered that 182 (91.96\%) out of 199 standard library modules are affected by these conflicts. The most frequently used standard library module names that conflict with third-party packages are \code{types}, \code{io}, and \code{logging}, which are used by 69,940, 47,214, and 35,694 packages, respectively. These results suggest that developers should be careful when choosing module names for their packages and avoid using names that already exist in the standard library.

\textbf{Findings.} We also observed a gap between the local development and the deployment environments that can lead to import confusion issues. When a program is developed locally, the current working directory has a higher priority than the standard library modules in \code{sys.path}. However, when the program is packaged and installed by others from PyPI, the \code{site-packages} directory has a lower priority than the standard library modules in \code{sys.path}. This gap can result in unexpected runtime errors due to importing wrong modules. To address this problem, developers have two options: changing the module name or using relative path import. However, the two solutions may break backward compatibility and reduce the readability or portability of the code.

In addition, we notice that the number of Module-to-Lib conflict packages increased each year. To further illustrate this trend, we plot the number and percentage of Module-to-Lib conflict packages for each year from 2005 to 2022. As shown in the \cref{fig: Libtimestamp}, both the number and percentage increased steadily over the years, indicating that this threat became more prevalent and severe as the Python ecosystem grew and the extensions to the standard library. This suggests that developers do not pay enough attention to the potential conflicts with the standard library modules when naming their modules, or they are unaware of the existing or newly added standard library modules that might conflict with their modules.

\textbf{Module-in-Dep conflict.}
We conducted an empirical study on the entire ecosystem of 4.2 million packages and detected 129,840 (3.07\%) packages with module-in-Dep conflicts, involving 11,666 projects. we also find 38,371 packages involving 4,516 projects that exhibit different module file contents but the same paths, which may cause functionality errors. Moreover, we noticed that some conflicting modules may change their contents after package updates, which could introduce new problems in the future. Although these conflicting modules may not be invoked at runtime, they do have the effect of module overwriting, which compromises the integrity of packages. There is also no guarantee that these modules will not be called and used in a future version.
% However, not all module-in-Dep conflicts have further impacts on the functionality of the packages. We identify three main reasons: (1) the overwritten module is not used at runtime (e.g., test module); (2) the overwritten module is an empty or identical file; (3) the two conflicting modules implement the same functionality and expose the same APIs. After filtering out these false positive cases, 

\textbf{Findings.} We further analyzed the characteristics of the conflicting packages. First, two packages from different maintainers that provide similar functionalities often use the same or similar module names. This is because they tend to copy from each other, thus avoiding unnecessary duplication of the wheel or convenient naming. Second, two packages related by migration often result in a conflict, where one package is deprecated and replaced by another package. Third, two packages that are different incompatible versions or variants of the same project. For example, in the dependency graph of \code{saleor}, \code{python-magic-bin} is a fork of \code{python-magic} with a different maintainer; in the dependency graph of \code{riffusion}, \code{soundfile} is a migrated version of \code{pysoundfile}; and \code{opencv-python} and \code{opencv-python-headless} are two distributions of \code{opencv} for different environments.

We also observed that these conflicts often occurred either in older continuous versions (2,342 out of 4,516) or in all versions (1,819 out of 4,516) of a project. This indicates that some developers or users discovered and resolved some conflicts when they encountered functionality issues, while others did not notice or update their dependencies. This implies that module-in-Dep conflicts have a certain persistence and concealment, which may affect the reliability of Python applications. For example, the project \code{aniposelib} used \code{opencv-python} and \code{opencv-contrib-python} dependencies prior to version 0.3.7, which was fixed by maintainers in a later version due to bugs raised in issue~\cite{aniposelib} caused by Module-in-Dep conflicts. What's more, such conflicts exist in an average of 6.5 versions. Such a large time gap can affect the functionality and maintainability of the project. Therefore, we argue that it is important to detect and prevent module-in-Dep conflicts in Python packages to ensure correct functionality.

From the time dimension, the number of packages with Module-in-Dep conflicts also gradually increases over time, as shown in \cref{fig: Libtimestamp}. Many older packages that didn't have conflicts before are coming back into conflict as dependencies are migrated. 
%https://github.com/lambdaloop/aniposelib/issues/2
% https://github.com/androguard/androguard/issues/236
% https://github.com/jupyter/jupyter_core/issues/55

\subsection{RQ3: GitHub Projects Study}
We select popular Python projects from GitHub. We collect the top 3,000 most-starred Python projects and 1,187 popular projects from awesome-Python\cite{pyawesome}. We merge and deduplicate the two datasets and obtain a total of 3,711 projects with 93,487 tags. We analyze their dependencies, resolve dependency graphs with \resolver, and detect module-in-Dep conflicts for them.

We detect 519 (13.93\%) projects with 10,850 (11.61\%) tags that have module overwriting threats. The results show that module-in-Dep conflicts are more prevalent in GitHub projects, as these projects tend to declare more dependencies than packages on PyPI. Although these conflicting modules may not affect the functionality of the program if the file contents are unchanged before and after overwriting, they can break the integrity of the package in the local environment and cause errors that are hard to debug. Moreover, there are 2,569 tags for 108 projects that may have functional errors, due to the difference in file contents before and after overwriting. Of the 108 projects, 65 are the latest version, while 43 projects have fixed module-in-Dep issues in later versions. This means that the module conflict problem is latent, with an average of 23 historical versions affected. It is often only when a user encounters an error that the maintainer becomes aware of the problem and fixes it. We manually analyze the conflicting modules in these 65 projects and report 35 issues to the project developers, of which 11 projects replied and 12 fixed the MC problems. The others do not respond, but since they have the same conflicting modules as the confirmed issues, we can assume that they have a real impact.

\textbf{Findings.}
We find that module conflicts occur more often in the AI field. This is because developers need to introduce one of the four \code{opencv-python} base packages when adding dependencies, along with other related AI projects. However, other related AI projects may also introduce other incompatible versions of the base package. These base packages are stated in the official documentation~\cite{opencv} that they cannot coexist because they all use the module name of \code{cv2}. This behavior is beyond the developer's control because they can only control direct dependencies, and the indirect dependencies are a black box. Such conflicts result in incompatibility between different AI projects when they are used together.

In addition, we find that some developers even include the same functional dependency in the direct dependency, and the two dependencies have module conflicts. Talking to the developers, they say that adding a dependency when they encountered an error could fix a strange error (which was actually caused by module overwriting). This means that developers tend to focus more on whether the program can run properly, and introduce functionally redundant dependencies, which not only increase the complexity of the project, but also increase the difficulty of building the project environment. To make matters worse, project issues reveal that many developers are not aware of module conflicts. They often add or remove dependencies after getting an error report to keep the program working. Of the 12 latest tags that were fixed, 10 were fixed by removing redundant dependencies. Therefore, our work reveals the nature and potential impact of module conflicts, and helps them to recognize and correctly declare dependencies to mitigate conflicts during the debugging phase.

\vspace{10pt}
\section{Limitations}
\label{sec:limitations}
\name can help developers detect potential module conflict threats during the development and packaging phase, thus helping them properly configure packaging scripts, but it faces the following threats.

First, in terms of detecting module-in-Dep conflicts, \name is responsible for detecting whether there are conflict modules in the dependencies declared by the project, which will definitely cause module overwriting and destroy the integrity of packages installed in the local environment. However, environmental damage does not necessarily affect the running of the program, because they may not have been imported or the files are empty (e.g., empty \code{__init__.py} file). For the sake of efficiency, our work did not take into account the call graph, which is time-consuming. But we will analyze it as our future work to optimize \name.

Second, when it comes to detecting module-to-Lib conflicts, \name doesn't know which standard library packages are cached by the user's native runtime. Because this involves the Python interpreter running time, Python version updates, etc., we have no way to collect relevant data from users due to privacy reasons. As a result, \name only collects and detects if the module names declared in packages conflict with the names in the standard library as a reminder, and we recommend that developers rename these modules that may conflict to prevent potential problems.

\section{Discussion}
\label{sec:disscussion}
We discuss from different perspectives: developers, PyPI maintainers, and future works.

\textbf{Developers.} For developers, when declaring dependencies, it is important to be aware of the potential module conflicts between dependencies, especially those that provide similar functionality. In addition, redundant dependencies should be avoided to prevent some errors caused by module overwriting during the installation process, which are often difficult to detect and debug. Furthermore, when developers declare a dependency, they should pay attention to whether the dependency is deprecated or migrated. Because these dependencies mean that there is no subsequent maintenance. If a vulnerability is disclosed, there is no maintainer to release a patch to fix it. We believe effective dependency management is very important, as it not only reduces the likelihood of module conflicts, but also ensures compatibility and functional integrity.

\textbf{PyPI maintainers.} Although Python provides the concept of namespace packages, the installation process for namespace packages is similar to that of regular packages and can still result in overwriting behavior. While the presence of \code{__init__.py} files in the same namespace does not cause impacts, because there is no real reason to call them, if there are other module conflicts between two packages within the same namespace, it can still lead to problems. In addition, we think \pip should isolate third-party packages from each other when installing them, like other languages do, so that different third-party packages cannot interact with each other. Moreover, \pip should warn about overwriting when appropriate, rather than overwriting by default, which can cause developers to destroy the local development environment without their notice. Python should provide a mechanism to select modules from third-party packages to be imported, instead of importing an entire module or importing some features from a module while missing the concept of packages.

\textbf{Future work.} We see two main directions for future work. First, existing developers lack analysis tools for their own dependencies and project profiles. This is mainly because we found that developers pack redundant files into packages and upload them to PyPI, or project developers declare redundant dependencies. Second, because the current tools of \name do not take into account the actual order of installation and whether the overwritten modules are actually imported, there is a certain false positive. However, these false positive examples do have overwriting problems, which we summarized in the issues study. Moreover, considering the efficiency of large-scale analysis, \name does not use the project's call graph to analyze whether the conflicting modules in Python projects have real import behavior. Therefore, we will make up for this deficiency in future work, so as to improve the accuracy of the tool.

\section{Related Work}
\label{sec:related-work}
% \textbf{Module conflict.}
The most relevant works to this paper are PyCRE~\cite{cheng2022conflict}, Pipreq~\cite{pipreq}, PyEGo~\cite{ye2022knowledge}, SnifferDog~\cite{wang2021restoring}, and PyDFix~\cite{mukherjee2021fixing}. 
PyCRE~\cite{cheng2022conflict} and PyDFix~\cite{mukherjee2021fixing} aim to solve the problem that some projects on PyPI cannot be installed properly due to environmental compatibility issues. PyCRE generates project dependencies through static analysis of calls between modules to form a domain knowledge graph, while PyDFix dynamically obtains the log content of installation failures to judge dependency errors and fixes them through continuous patching. However, our work focuses on the module conflicts that occur during and after the installation of PyPI projects, assuming that the installation process is successful. In addition, in terms of module extraction, PyCRE not only does not notice that the module path will change before and after package installation, but also does not notice that modules with the same import statement may belong to different packages.

Pipreq is a tool that generates the \code{requirement.txt} from the project source code. It treats modules and packages as one-to-one mappings, and doesn't deal with the case where different packages can contain the same module path.
PyEGo~\cite{ye2022knowledge} is a tool that automatically infers the dependencies of third-party packages, the Python interpreter, and system libraries at compatible versions for Python programs. It only extracts module paths from the metadata file. For cases where different packages can contain the same module path, it considers the conflicting module to belong to the most popular package.
Vu et al.~\cite{Vu_2020} study the possible attack vectors in the Python ecosystem, but do not delve into the MC problem. 
SnifferDog~\cite{wang2021restoring} fixes the environment of the Jupyter Notebook project using module information and dependency information. It also uses a one-to-one mapping, so it doesn't notice the module conflict problem either.

On the contrary, our work mainly analyzes the impact of MC problem in the ecosystem on a large scale, which is not involved in the above work. They either did not cover the module extraction process, or simply took the module information from the file (low accuracy), and did not pay attention to the complex problem of mapping the module before and after installation. For the dependency resolution part, they use their own specific algorithm for resolution, but the resolution rules are different from the resolution rules used in the actual \pip installation. In general, we noticed some aspects in the module aspect that others had not noticed before, and used \pip-compatible algorithms in dependency resolution with higher accuracy.

% \textbf{Dependency resolution.} Many related works rely on dependency resolution as part of their work, but the accuracy of their tools is not well assessed. 
\section{Conclusion}
\label{sec:conclusion}
This paper makes a systematic empirical study of module conflicts in Python. We implemented a tool called \name. It parses module information and dependency information from the PyPI project and, in turn, detects three types of module conflicts. We used \name to detect 4.2 million version packages and 3,711 popular GitHub projects. We identified 108 GitHub projects with module-in-Dep conflicts and reported issues to them and we get 12 fixed and good feedback. All experimental data in this paper are available at \url{https://sites.google.com/view/moduleguard}.

% \balance

\begin{acks}
% \section*{Acknowledgments}
% \wenbo{format need to change}
The authors would like to thank all reviewers sincerely for their valuable comments. This work is partially supported by the National Key R\&D Program of China (2022YFB3103900). It is also supported by the National Research Foundation, Singapore, and DSO National Laboratories under the AI Singapore Programme (AISG Award No: AISG2-GC-2023-008). It is also supported by the National Research Foundation, Singapore, and the Cyber Security Agency under its National Cybersecurity R\&D Programme (NCRP25-P04-TAICeN) and the NRF Investigatorship NRF-NRFI06-2020-0001. Any opinions, findings, conclusions, or recommendations expressed in this material are those of the author(s) and do not reflect the views of the National Research Foundation, Singapore, and the Cyber Security Agency of Singapore.

\end{acks}

% \clearpage
% \balance
\bibliographystyle{ACM_Reference_Format}
\balance
\bibliography{ref}

\end{document}